# Achieve Better Ranking Accuracy Using CloudRank Framework for Cloud Services

Ms. M. Subha[#1], Mr. K. Saravanan[*2]

[#] Student, [*]Assistant Professor
Department of Computer Science and Engineering
Regional Centre of Anna University
Tirunelveli (T.N) India

***Abstract*— Building high quality cloud applications becomes an urgently required research problem. Nonfunctional performance of cloud services is usually described by quality-of-service (QoS). In cloud applications, cloud services are invoked remotely by internet connections. The QoS Ranking of cloud services for a user cannot be transferred directly to another user, since the locations of the cloud applications are quite different. Personalized QoS Ranking is required to evaluate all candidate services at the user - side but it is impractical in reality. To get QoS values, the service candidates are usually required and it's very expensive. To avoid time consuming and expensive real-world service invocations, this paper proposes a CloudRank framework which predicts the QoS ranking directly without predicting the corresponding QoS values. This framework provides an accurate ranking but the QoS values are same in both algorithms so, an optimal VM allocation policy is used to improve the QoS performance of cloud services and it also provides better ranking accuracy than CloudRank2 algorithm.**

## I. Introduction

Cloud computing can be defined as "a type of parallel and distributed system consisting of a collection of interconnected and virtualized computers that are dynamically provisioned and presented as one or more unified computing resources based on service-level agreements established through negotiation between the service provider and consumers" The consumer of the cloud can obtain the services through the network. In other words, users are using or buying computing services from others. Cloud can provide Anything as a Service (AaaS). In general, cloud provides application, computation power, storage, bandwidth, database etc.

The cloud removes the need for you to be in the same physical location as the hardware that stores your data. There are number of functionally equivalent services in the cloud Due to unreliable internet connections different cloud applications may receive different levels of quality for same cloud services so that optimal service selection becomes important. Cloud computing provides three main services, namely Software as a Service (SaaS), Platform as a Service (PaaS), and Infrastructure as a Service (IaaS). In Software as a Service (SaaS), Clients can use the software to provide by the provider, which usually need not to install and it is usually a one of many services. Like Gmail, search engine. In Platform as a Service (PaaS), Clients can run their own applications on the platform provided; General platforms are Linux and Windows. In Infrastructure as a Service (IaaS), Client can put their own operating system on cloud.

### A. Optimal Service Selection

QoS is an important research topic in cloud computing. Since there are a number of functionally equivalent services in the cloud, optimal service selection becomes important. When making an optimal cloud service selection from a set of functionally equivalent services, QoS values of cloud services provide valuable information to assist decision making. Client-side performance of cloud services is thus greatly influenced by the unreliable internet connections. Therefore, different cloud applications may receive different levels of quality for the same cloud service. The training data in the CloudRank framework can be obtained from the QoS values collected by monitoring cloud services.

The most straightforward approach of personalized cloud service QoS ranking is to evaluate all the candidate services at the user-side and rank the services based on the observed QoS values. However, this approach is impractical in reality, since invocations of cloud services may be charged. It is difficult for the cloud application designers to evaluate all the cloud services efficiently. To attack this critical challenge, we propose a personalized ranking prediction framework, named CloudRank, to predict the QoS ranking of a set of cloud services without requiring additional real-world service invocations from the intended users. The objective of Cloud Service Ranking is to build High-quality cloud applications for cloud services by using the QoS Ranking prediction Framework. CloudRank2_modify algorithm is used for optimal VM allocation for each service. Features of this paper as follows,

- It identifies the critical problem of personalized QoS ranking for cloud services and proposes a QoS ranking prediction framework to address the problem. To the best of our knowledge, CloudRank is the first personalized QoS ranking prediction framework for cloud services.

- It takes the advantage of the past usage experiences of other users for making personalized ranking predictions for the current user.
- It provides optimal VM allocation for each service which is used by service users.

## II. RELATED WORK

### A. QoS Ranking Prediction on Cloud Services

Since this work explores the issue of building high quality cloud applications. Quality-of-Service (QoS) is usually employed for describing the non-functional characteristics of Web services and employed as an important differentiating point of different Web services. Users in different geographic locations collaborate with each other to evaluate the target Web services and share their observed Web service QoS information.

In optimal Service Selection [2] proposed Exact and approximated algorithms for optimal service selection based on a given set of service requests (such as the activities occurring in a workflow), a set of service users (the available services), the result of the matchmaking process (that associates each request to the set of users that can satisfy it), and a numeric preference measure. It identified the Service Selection Problem (SSP). We show that the high computational complexity of the service selection problem is caused by the one-time costs associated with service users (e.g., Initialization and registration costs). In the absence of one-time costs, the optimal selection problem can be solved in polynomial time by applying a greedy approach. The heuristic algorithm seems to be faster, but it has no guarantees on the quality of the solution.

Collaborative filtering algorithms [3] proposed Memory-based algorithm and Model-based algorithm that predicts the utility of items to a particular user (the active user) based on a database of user votes from a sample or population of other users (the user database). We use two basic classes of evaluation metrics. The first characterizes accuracy over a set of individual predictions in terms of average absolute deviation. The second estimates the utility of a ranked list of suggested items. Bayesian networks typically have smaller memory requirements and allow for faster predictions than a memory-based technique such as correlation but Bayesian methods examined here require a learning phase that can take up to several hours and results are reflected in the recommendations.

Item-Based Top-*N* Recommendation Algorithms [4] determines the similarities between the various items from the set of items to be recommended. The key steps in this class of algorithms are (i) the method used to compute the similarity between the items, and (ii) the method used to combine these similarities in order to compute the similarity between a basket of items and a candidate recommender item. The goal of top-N recommendation algorithm was to classify the items purchased by an individual user into two classes: like and dislike. This algorithm is faster than the traditional user-neighborhood based recommender systems and it provides recommendations with comparable or better quality. The proposed algorithms are independent of the size of the user–item matrix.

Recommendation Algorithm [6] determines a set of customers whose purchased and rated items overlap the user's purchased and rated items. The algorithm aggregates items from these similar customers, eliminates items the user has already purchased or rated, and recommends the remaining items to the user. It generates high quality recommendations and the algorithm must respond immediately to new information. It is used to personalize the online store for each customer but it needs to apply recommendation algorithms for targeted marketing, both online and offline.

Collaborative filtering approach [7] addresses the item ranking problem directly by modelling user preferences derived from the ratings. It performs ranking items based on the preferences of similar users.

CloudRank approach [11] proposed greedy algorithm and it rank the component instead of service but this algorithm. It is used to rank a set of items, which treats the explicitly rated items and the unrated items equally. It does not guarantee that the explicitly rated items will be ranked correctly.

QoS-Aware Web Service by Collaborative Filtering [12] proposed Hybrid collaborative filtering method that To improve performance of Recommender System. It includes a user-contribution mechanism for Web service QoS information collection and an effective and novel hybrid collaborative filtering algorithm for Web service QoS value prediction. It is used to collect systematic QoS information and it provides better feasibility of WSRec (Web service recommender system) but it needs to monitor more real-world Web services and it needs to investigate more QoS properties of Web services.

## III. PROPOSED SYSTEM

Quality-of-service can be measured either at the server side or at the client side. Client-side QoS properties provide more accurate measurements of the user usage experience. The commonly used client-side QoS properties include response time, throughput, failure probability, etc. This paper mainly focuses on ranking prediction of client-side QoS properties, which likely have different values for different users (or user applications) of the same cloud service.

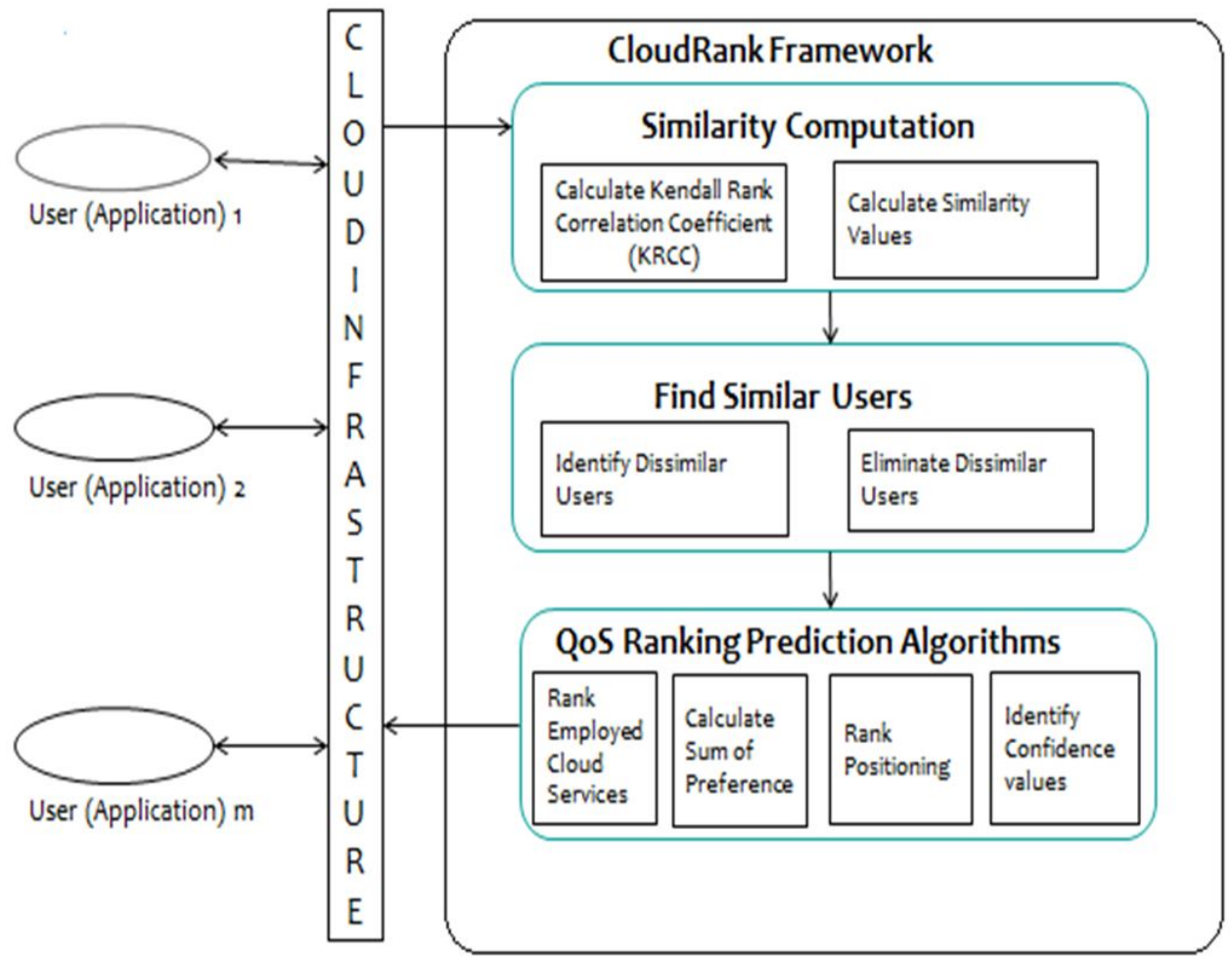

Fig. 1 System Architecture of CloudRank

*A. CloudRank2_modify Algorithm*

In the existing system, ranking is implemented based on preference values and Confidence values of cloud services. CloudRank2 achieve better ranking than CloudRank1 using confidence values of cloud services but both algorithms produce same QoS performance which is shown in figure 8 & 9. To get better ranking accuracy and high quality if cloud application, this paper proposed CloudRank2_modify algorithm.

*1) Find Similar Users:*

The similarity between active users and training users are calculated using Kendall Rank Correlation Coefficient (KRCC). It evaluates the degree of similarity by considering the number of inversions of service pairs which would be needed to transform one rank order into the other. The KRCC value of users u and v can be calculated by,

$$Sim(u, v) = \frac{C-D}{N(N-1)/\,2} \qquad (1)$$

Where N is the number of services, C is the number of concordant between two lists, D is the number of discordant pairs, and there is totally N (N-1) /2 pairs for N cloud services.

*2) Eliminate Dissimilar Users:*

The similar users can be identified by calculating similarity values. However, employing QoS values of dissimilar users will greatly influence the prediction accuracy. Our previous approaches include dissimilar users. We exclude the users with negative correlations (negative similarity values) and only employ the Top-K similar users for making QoS ranking predictions. By eliminating the dissimilar users, we get correct prediction accuracy. In our approach, a set of similar users is identified for the active user u by

$$N(u) = \{v | v \in T_u, Sim(u, v) > 0, v \neq u\} \qquad (2)$$

Where $T_u$ is a set of the Top-K similar users to the user u and Sim (u, v) >0 excludes the dissimilar users with negative similarity values. The value of Sim (u, v) in 2 is calculated by equation (1).

*3) Calculate Preference Values:*

Calculate the sum of preference values with all other services by $\pi(i) = \sum_{j \in I} \psi(i, j)$. Larger π(i) value indicates more services are less than i. The value of the preference function $\psi$ (i, j) is anti-symmetric, i.e., $\psi(i.j)$= - $\psi(j, i)$. The preference function $\psi$ (i, j) where service i and service j are not explicitly observed by the current user u.

$$\psi(i, j) = \sum_{v \in N(u)^{ij}} w_v \ (q_{v,i} - q_{v,j}) \qquad (3)$$

$w_v$ is a weighting factor of the similar user v, which can be calculated by

$$w_v = \frac{Sim(u,v)}{\sum_{v \in N(u)^{ij}} Sim(u,v)} \qquad (4)$$

$w_v$ makes sure that a similar user with higher similarity value has greater impact on the preference value prediction in (3). With (3) and (4), the preference value between a pair of services can be obtained by taking advantage of the past usage experiences of similar users.

*4) Calculate Confidence Values:*

The preference values $\psi(i, j)$ in the CloudRank1 algorithm can be obtained explicitly or implicitly. When the active user has QoS values on both the services i and service j, the preference value is obtained explicitly. Assuming there are three cloud services a, b, and c. The active users have invoked service a and service b previously. The list below shows how the preference values of $\psi(a, b), \psi(a, c), and\ \psi\ (b, c)$ can be obtained explicitly or implicitly.

- $\psi(a, b)$ : obtained explicitly.
- $\psi(a, c)$ : obtained implicitly by similar users with similarities of 0.1, 0.2, and 0.3.
- $\psi(b, c)$: obtained implicitly by similar users with similarities of 0.7, 0.8, and 0.9.

In the above example, we can see that different preference values have different confidence levels. It is clear that $C(a, b) > C(b, c) > C(a, c)$ where C represents the confidence values of different preference values. The confidence value of $\psi(a,b)$ is higher than $\psi(a,c)$ , since the similar users of $\psi(b,c)$ have higher similarity. CloudRank2, which uses the following, rules to calculate the confidence values. If the user has QoS value of these two services i and j. The confidence of the preference value is 1. When employing similar users for the preference value prediction, the confidence is determined by similarities of similar users as follows,

$$C(i,j) = \sum_{v\epsilon N(u)^{ij}} w_v \, Sim(u,v) \qquad (5)$$

Where v is a similar user of the current active user u, $N(u)^{ij}$ is a subset of similar users, who obtain QoS values of both services i and j, and $w_v$ is a weighting factor of the similar user v, which can be calculated by (4).

$w_v$ makes sure that a similar user with higher similarity value has greater impact on the confidence calculation. Equation (4) guarantees that similar users with higher similarities will generate higher confidence values. This algorithm achieved more accurate ranking prediction of cloud services. An optimal VM allocation is implemented with CloudRank2 algorithm.

5) *Optimal VM Allocation:*

The optimal VM allocation approach is used to allocate each service to the corresponding Virtual Machine. First, the VM details of each service is calculated. Based on the details, the cloud services are allocated to each Virtual Machine. It is implemented using CloudSim. Finally, It provides fast response for each service and it also provides high throughput. CloudRank2_modify algorithm provides high QoS performance and also it provides better ranking accuracy than existing algorithms.

## IV. Simulation and Results

The CloudSim simulation layer provides support for modelling and simulation of virtualized Cloud-based datacenter environments including dedicated management interfaces for virtual machines (VMs), memory, storage, and bandwidth. The fundamental issues such as provisioning of hosts to VMs, managing application execution, and monitoring dynamic system state are handled by this layer.

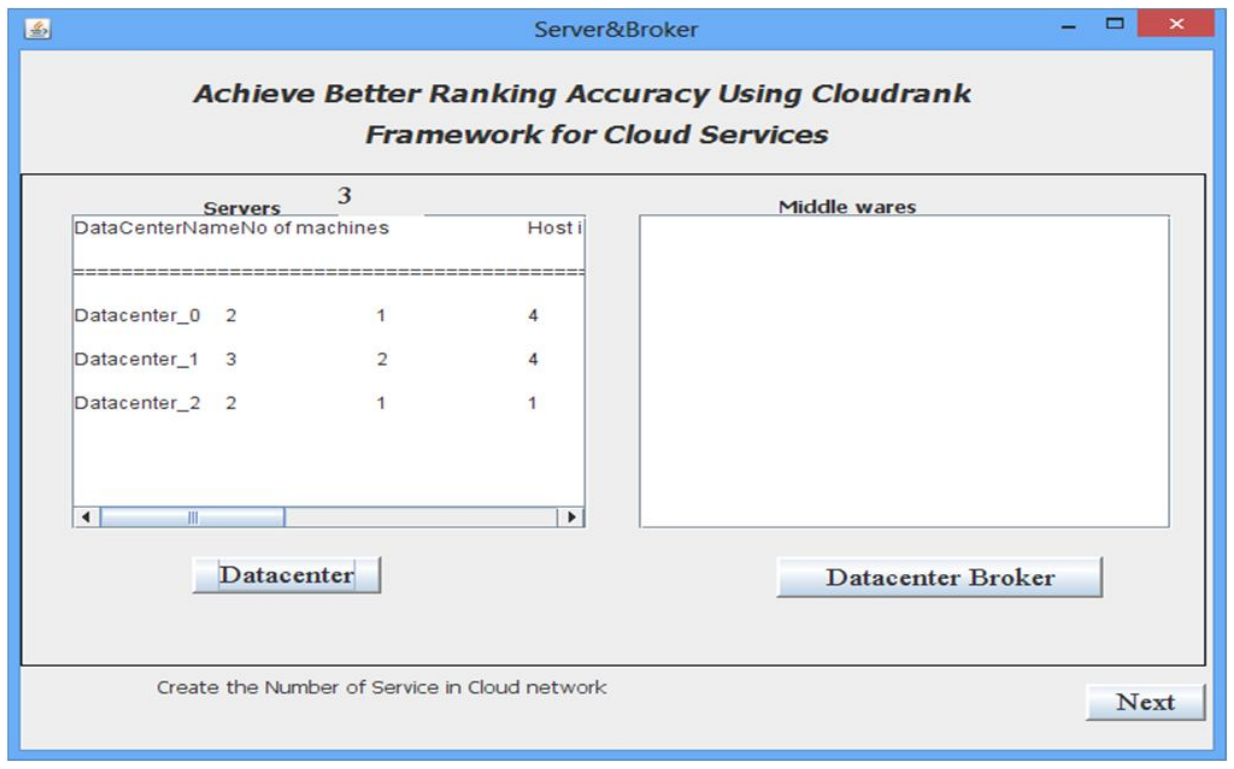

Fig. 2 Datacenter Creation

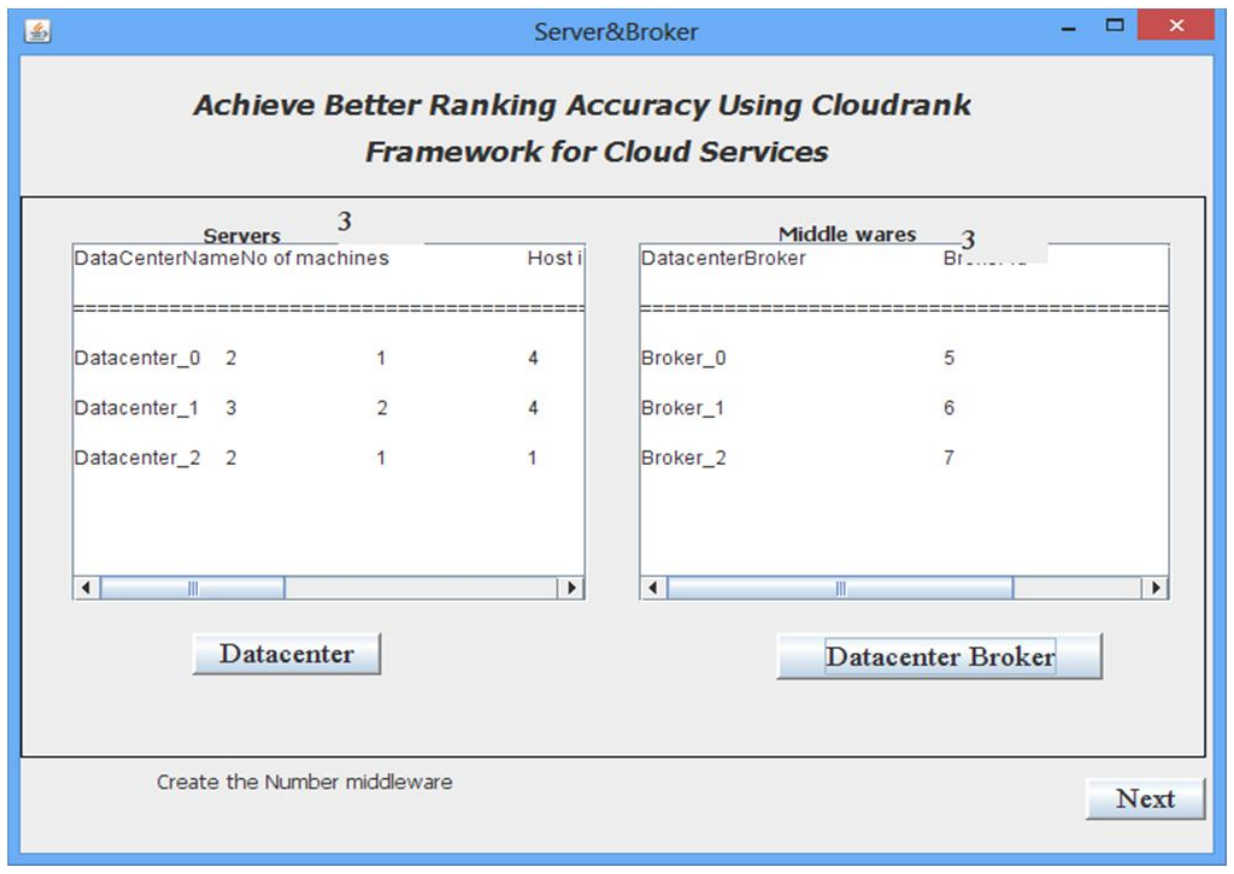

Fig. 3 DatacenterBroker Creation

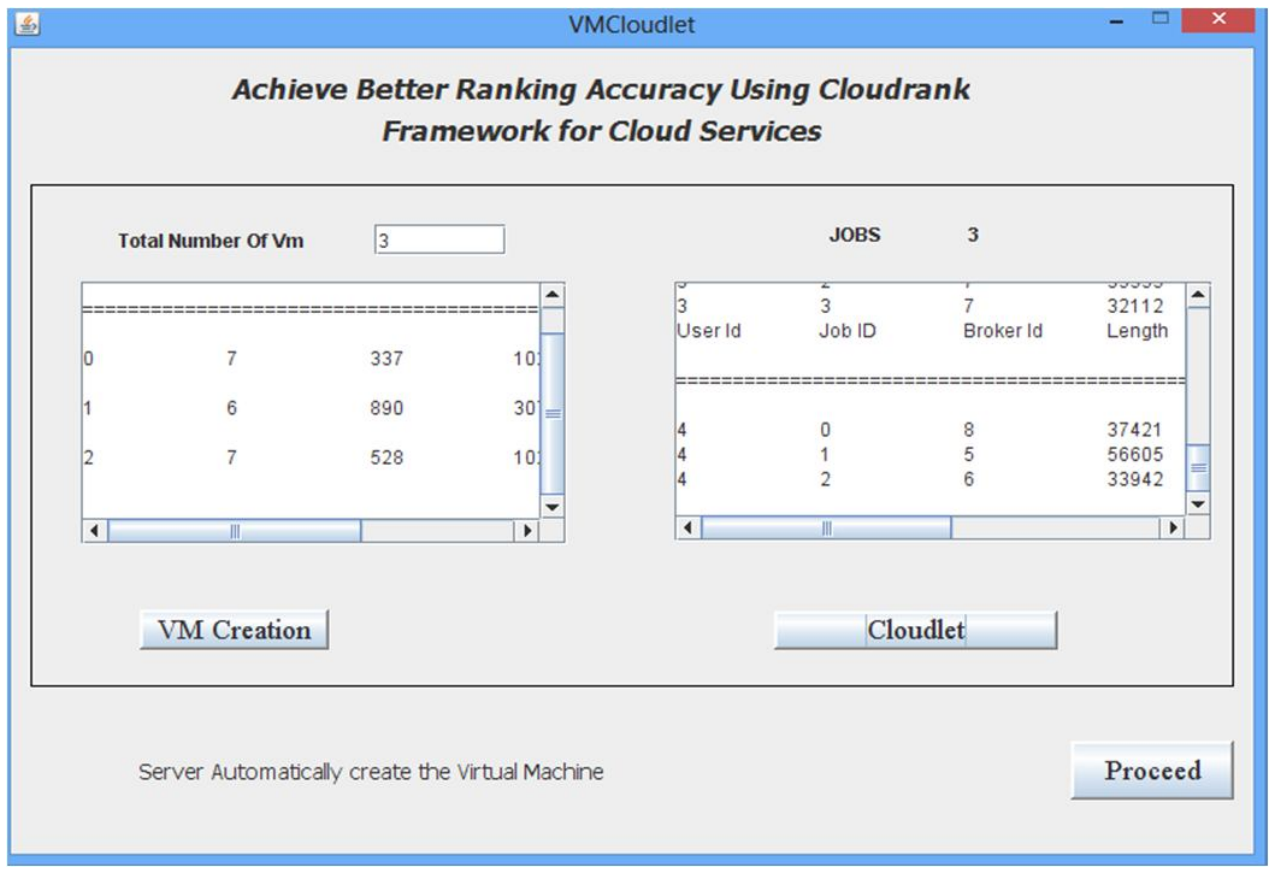

Fig. 4 Cloudlet Creation

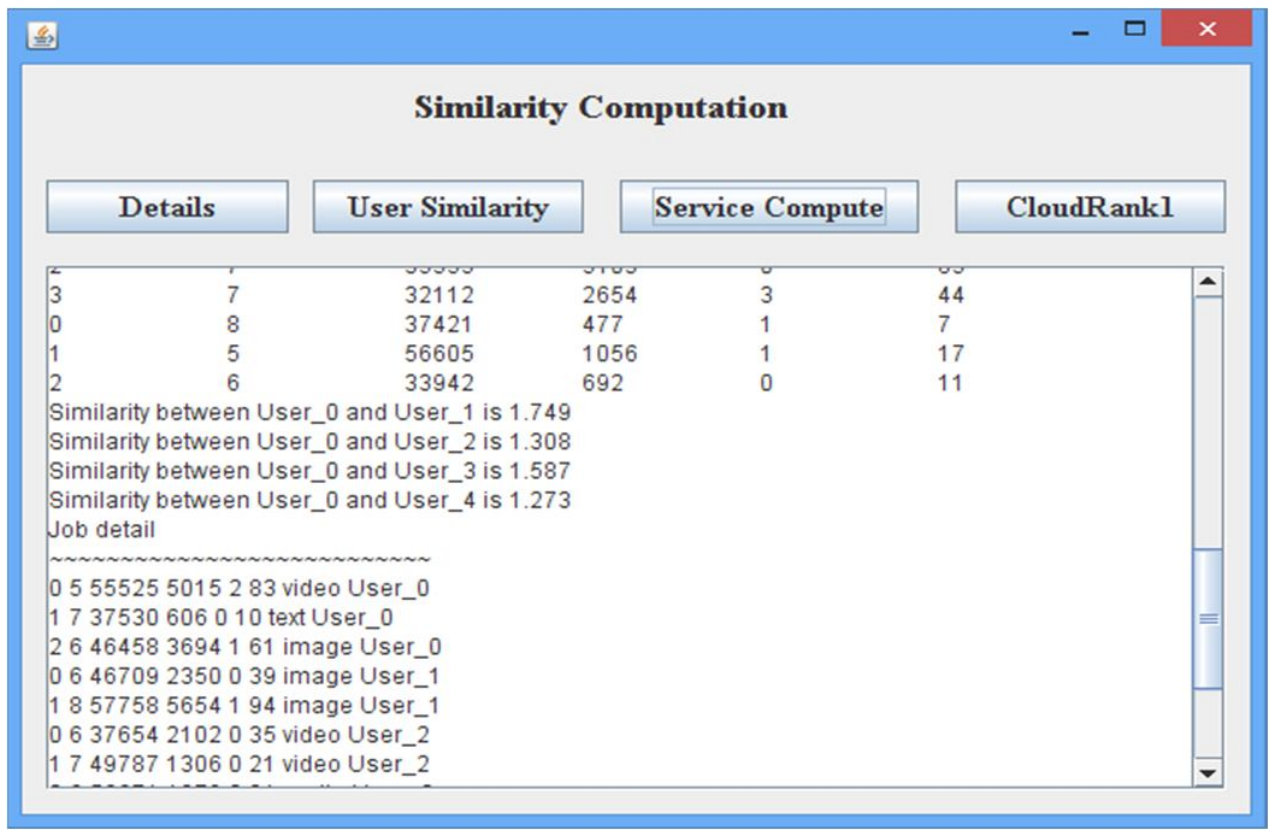

Fig. 5 Similarity Computation

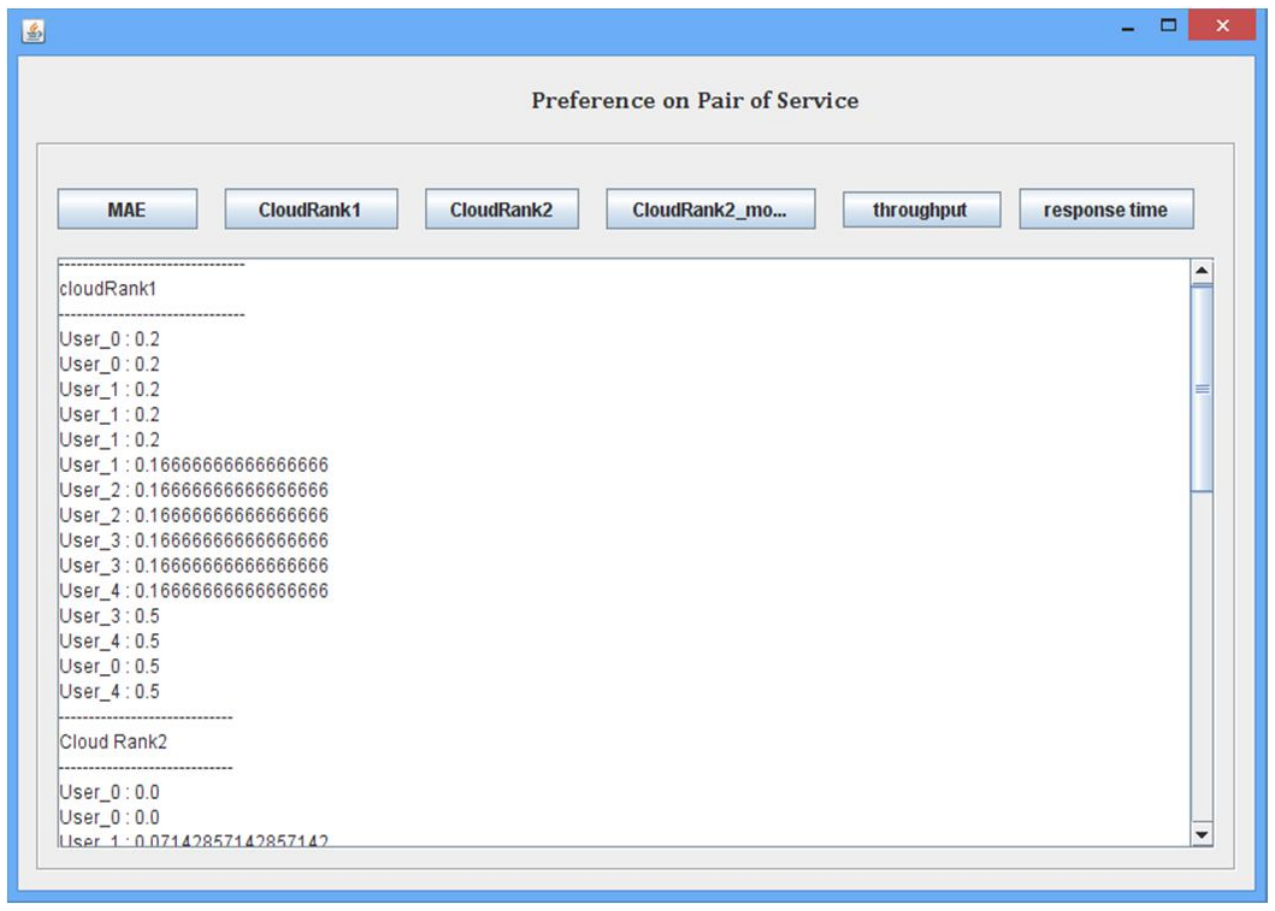

Fig. 6 CloudRank1 and CloudRank2 Algorithms

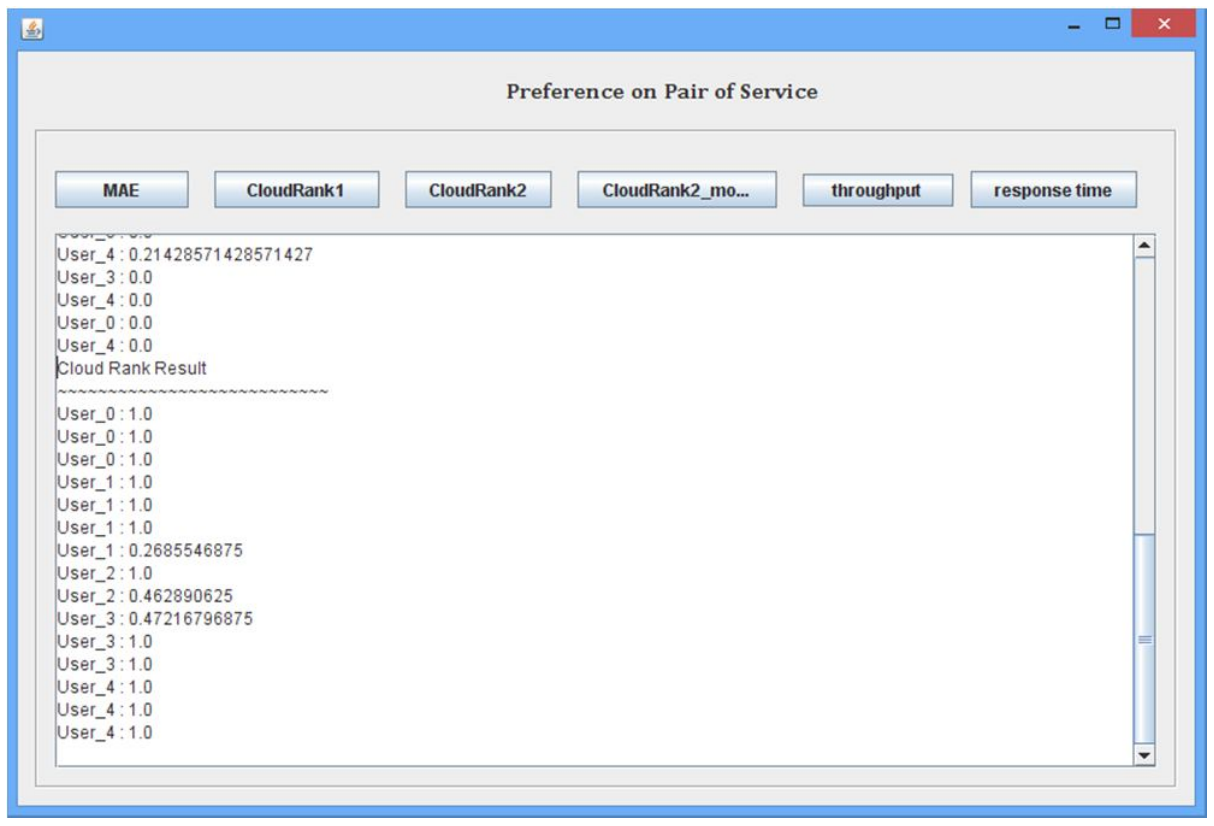

Fig. 7 CloudRank2_modify Algorithm

Throughput represents the data transfer rate over the network. In Figure the throughput for all three algorithms are calculated. CloudRank1 and CloudRank2 are having same throughput values but their ranking is different.

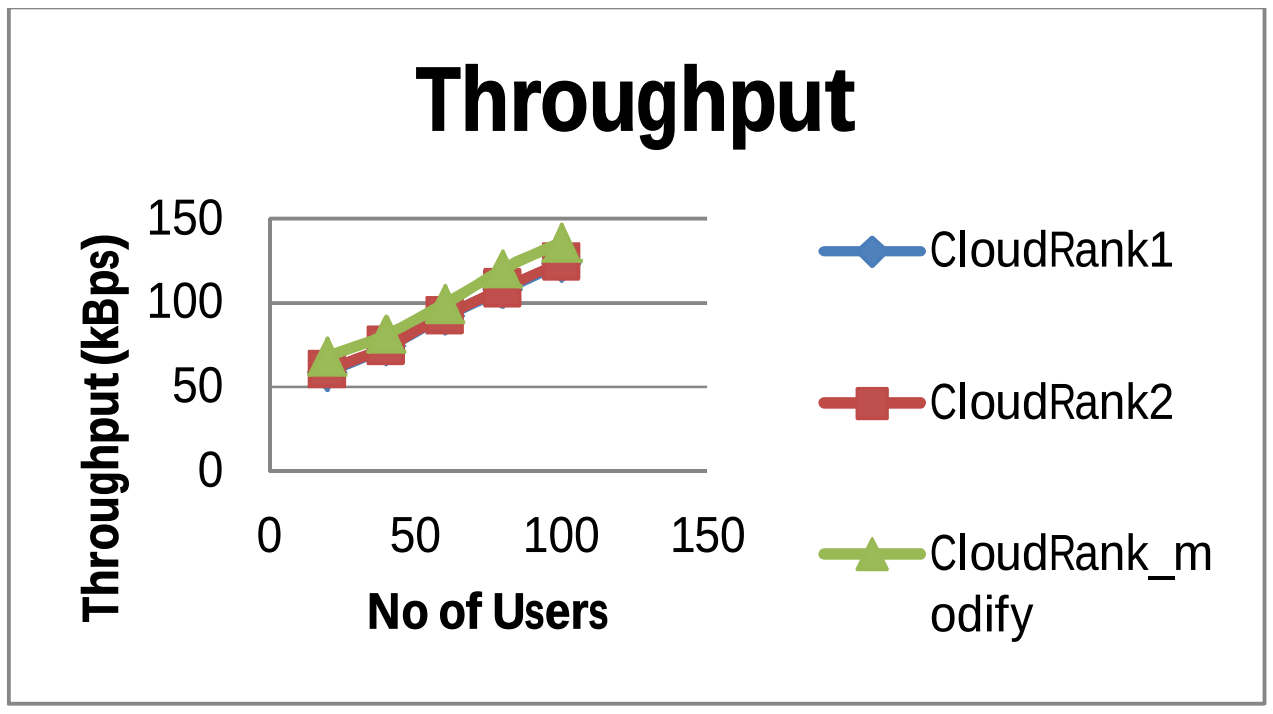

Fig. 8 Throughput Computation

Response-time refers to the time duration between the users sending out a request to a service and receiving a response. CloudRank1 and CloudRank2 are having same response-time values but their ranking is different.

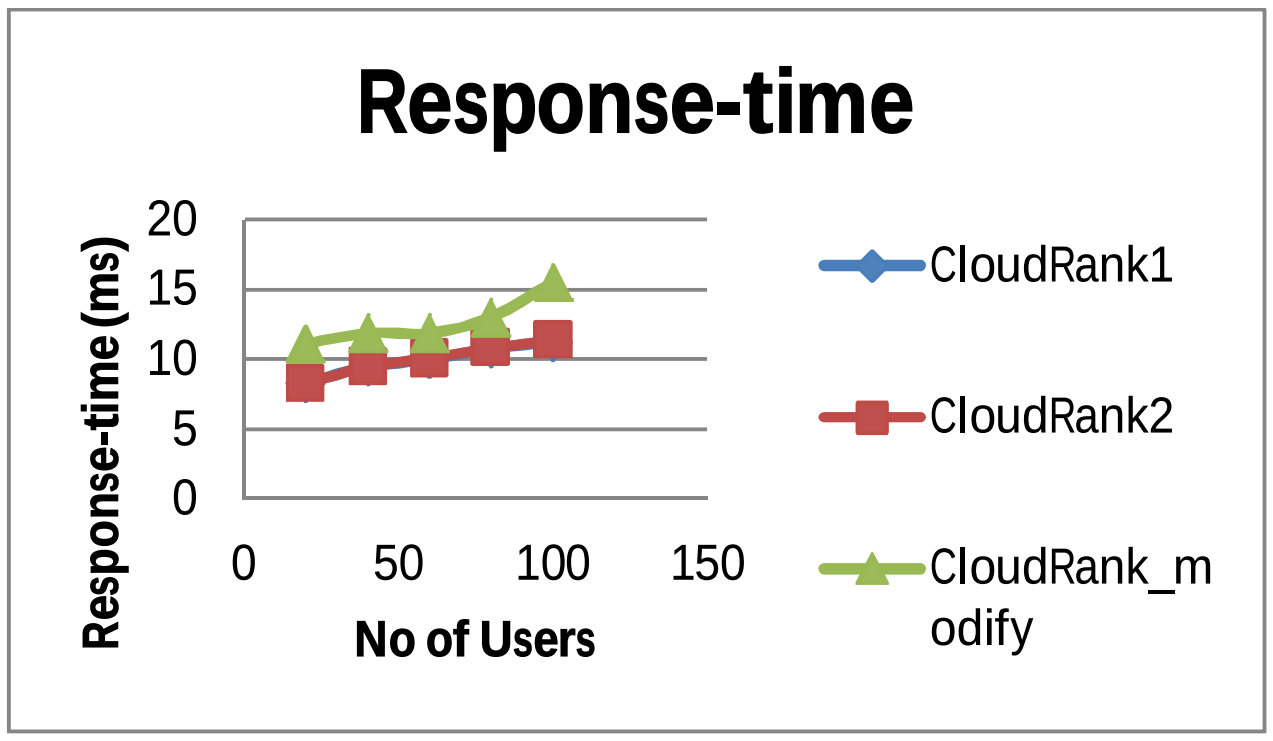

Fig. 9 Response-time Computation

In this QoS Ranking, the efficiency is calculated based on response-time and throughput for each service. In this figure shows that the CloudRank2_modify algorithm provides better ranking accuracy than the other two algorithms.

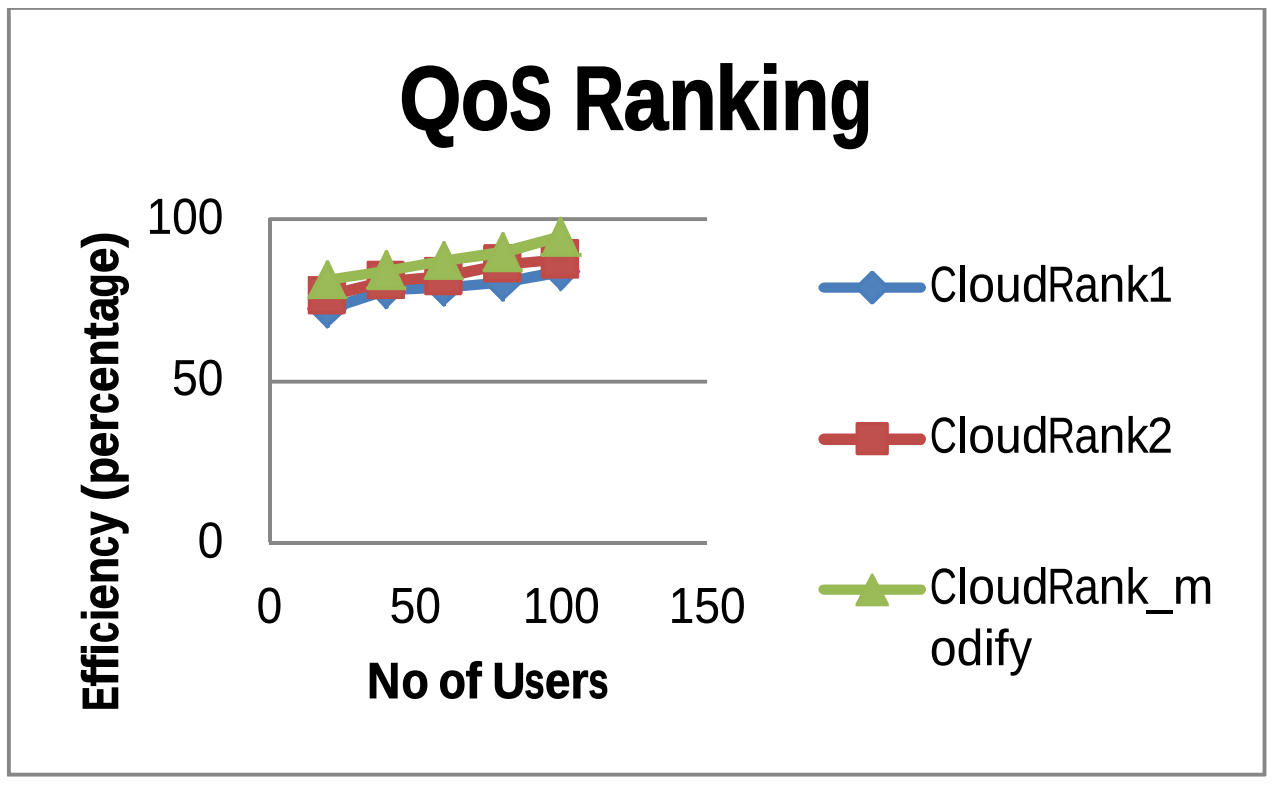

Fig. 10 Ranking Computation

## V. CONCLUSIONS

In this paper, the three Ranking Prediction schemes, CloudRank1, CloudRank2 and CloudRank2_modify are investigated and their performance is compared. The cloud environment is created and the CloudRank algorithm is implemented using CloudSim. The QoS performance and Ranking are shown in figure 8, 9 & 10. CloudRank2_modify algorithm provides optimal VM allocation to each service so that it increases the QoS performance and better ranking than the other two algorithms. Future research will focus on investigating and improving ranking accuracy of our approaches by using various ranking techniques.